\documentclass[%
 reprint,
superscriptaddress,
groupedaddress,
nofootinbib,
 amsmath,amssymb,
 aps,
floatfix,
]{revtex4-2}

\usepackage{graphicx}
\usepackage{dcolumn}
\usepackage{bm}
\usepackage{hyperref}

\usepackage{xspace}
\usepackage{xcolor}
\usepackage{booktabs}
\usepackage{makecell}
\newtheorem{property}{Property}

\newcommand{\kt}{\ensuremath{k_{\mathrm{T}}}\xspace}
\newcommand{\pt}{\ensuremath{p_{\mathrm{T}}}\xspace}
\newcommand{\msd}{\ensuremath{m_{\mathrm{SD}}}\xspace}
\newcommand{\mjj}{\ensuremath{m_{JJ}}\xspace}
\newcommand{\invfb}{\ensuremath{\,\text{fb}^{-1}}\xspace}
\newcommand{\astinpar}{{\ensuremath{\scalebox{0.7}{(}*\scalebox{0.7}{)}}}}
\newcommand{\MADGRAPH} {\textsc{MadGraph}\xspace}
\newcommand{\MCATNLO} {\textsc{mc@nlo}\xspace}
\newcommand{\MGvATNLO}{\MADGRAPH{}5\_a\MCATNLO}
\newcommand{\PYTHIA} {{\textsc{pythia}}\xspace}
\newcommand{\DELPHES} {{\textsc{delphes}}\xspace}
\newcommand{\jetclass} {{\textsc{JetClass}}\xspace}
\newcommand{\jetclassii} {{\textsc{JetClass-II}}\xspace}
\newcommand{\xbb}{\ensuremath{X\to bb}\xspace}
\newcommand{\xbs}{\ensuremath{X\to bs}\xspace}
\newcommand{\xbq}{\ensuremath{X\to bq}\xspace}
\newcommand{\xqq}{\ensuremath{X\to qq}\xspace}
\newcommand{\xcs}{\ensuremath{X\to cs}\xspace}

\begin{document}

\preprint{APS/123-QED}

\title{Accelerating Resonance Searches via Signature-Oriented Pre-training}

\author{Congqiao Li}
\email{congqiao.li@cern.ch}
\affiliation{School of Physics and State Key Laboratory of Nuclear Physics and Technology, Peking University, 100871 Beijing, China}

\author{Antonios Agapitos}
\affiliation{School of Physics and State Key Laboratory of Nuclear Physics and Technology, Peking University, 100871 Beijing, China}

\author{Jovin Drews}
\affiliation{Institute for Experimental Physics, Universit\"at Hamburg, Luruper Chaussee 149, 22761 Hamburg, Germany}

\author{Javier Duarte}
\affiliation{University of California San Diego, La Jolla, CA 92093, USA}

\author{Dawei Fu}
\affiliation{School of Physics and State Key Laboratory of Nuclear Physics and Technology, Peking University, 100871 Beijing, China}

\author{Leyun Gao}
\affiliation{School of Physics and State Key Laboratory of Nuclear Physics and Technology, Peking University, 100871 Beijing, China}

\author{Raghav Kansal}
\affiliation{University of California San Diego, La Jolla, CA 92093, USA}

\author{Gregor Kasieczka}
\affiliation{Institute for Experimental Physics, Universit\"at Hamburg, Luruper Chaussee 149, 22761 Hamburg, Germany}

\author{Louis Moureaux}
\affiliation{Institute for Experimental Physics, Universit\"at Hamburg, Luruper Chaussee 149, 22761 Hamburg, Germany}

\author{Huilin Qu}
\affiliation{CERN, EP Department, CH-1121 Geneva 23, Switzerland}

\author{Cristina Mantilla Suarez}
\affiliation{Particle Physics Division, Fermi National Accelerator Laboratory, Batavia, IL 60510, USA}

\author{Qiang Li}
\email{qliphy0@pku.edu.cn}
\affiliation{School of Physics and State Key Laboratory of Nuclear Physics and Technology, Peking University, 100871 Beijing, China}

\date{\today}

\begin{abstract}
The search for heavy resonances beyond the Standard Model (BSM) is a key objective at the LHC. 
While the recent use of advanced deep neural networks for boosted-jet tagging significantly enhances the sensitivity of dedicated searches, it is limited to specific final states, leaving vast potential BSM phase space underexplored.
We introduce a novel experimental method, Signature-Oriented Pre-training for Heavy-resonance ObservatioN (Sophon), which leverages deep learning to cover an extensive number of boosted final states.
Pre-trained on the comprehensive \textsc{JetClass-II} dataset, the Sophon model learns intricate jet signatures, ensuring the optimal constructions of various jet tagging discriminates and enabling high-performance transfer learning capabilities.
We show that the method can not only push widespread model-specific searches to their sensitivity frontier, but also greatly improve model-agnostic approaches, accelerating LHC resonance searches in a broad sense.
\end{abstract}

\maketitle


\section{Introduction}

Discovery of heavy resonances beyond the Standard Model (BSM) is a long-standing goal of the LHC program.
Despite tremendous efforts to search for resonances up to the TeV mass scale, no concrete evidence of a BSM resonance has been established~\cite{ATLAS:2024lda,atlasexoticstwiki,atlassusytwiki,atlashdbspublictwiki,cmsexoticstwiki,cmssusytwiki,cmsb2gtwiki}.
To date, besides these extensive experiments focusing on specific theoretical models, model-agnostic search techniques have also seen consistent progress~\cite{Collins:2018epr,Nachman:2020lpy,Andreassen:2020nkr,Stein:2020rou,Amram:2020ykb,Park:2020pak,Kasieczka:2021xcg,Tsan:2021brw,Hallin:2021wme,Hallin:2022eoq,Raine:2022hht,Farina:2018fyg,Heimel:2018mkt,Collins:2021nxn,Belis:2023mqs}
and their experimental implementations have been initiated~\cite{ATLAS:2020iwa,ATLAS:2023ixc,CMS-PAS-EXO-22-026}.
Their common goal is to enhance the sensitivity to new physics as much as possible in potentially unexpected phase space.

The boosted topology is widely explored in BSM searches at the ATLAS and CMS experiments 
as it focuses on high-momentum phase space where high-mass-scale new physics is likely to appear first.
When probing signals with boosted hadronic final states, recent LHC measurements of Higgs boson properties~\cite{CMS:HbbFullRun2,CMS:VHcc,CMS:ggHcc,CMS:HH4b} reveal that the main driver of the sensitivity is the enhanced performance of the deep neural network (DNN) used for large-radius (large-$R$) jet tagging~\cite{CMS-DP-2020-002,CMS-DP-2022-041,ATL-PHYS-PUB-2023-021} resulting from rapid progress in deep learning applied to jet tagging~\cite{Kasieczka:2019dbj,Moreno:2019bmu,Moreno:2019neq,pmlr-v162-qu22b}.
Training and deploying state-of-the-art jet networks in all possible boosted-jet final states should bring us to the sensitivity frontier for various BSM signal searches.
However, current boosted-jet taggers deployed in experiments cover limited final states as they are developed for specific tagging purposes~\cite{CMS-DP-2020-002,CMS-DP-2022-041,ATL-PHYS-PUB-2023-021,CMS:JMETagger,CMS:BTVFlvTagger,ATLAS:bbTaggerCalibPaper,ATLAS:TopTagger,ATLAS:WZTagger}.
In contrast, unknown BSM processes may produce jets with unpredictable signatures and may be initiated from arbitrary combinations of SM particles~\cite{Craig:2016rqv,Kim:2019rhy}.
This leaves the majority of such BSM signal phase space underexplored.
Therefore, a tool that enables us to push a broad range of final states towards their sensitivity frontier will accelerate our search for heavy new resonances at the LHC.

\begin{figure}[htbp]
\begin{center}
\includegraphics[width=0.48\textwidth]{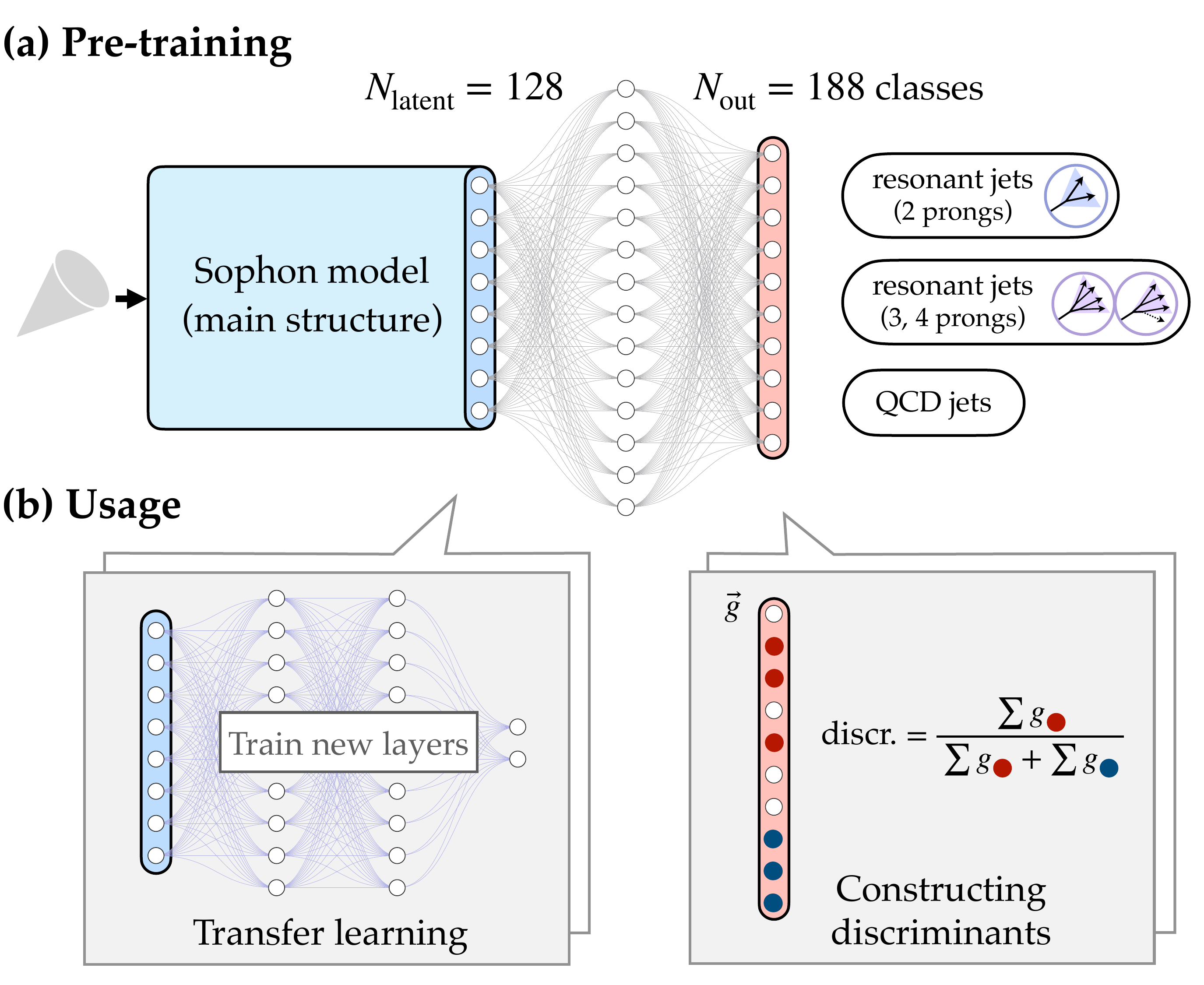}
\caption{Illustration of (a) the Sophon model pre-trained as a large-scale classifier over 188 classes of finely categorized jet signatures, and (b) the usage of the Sophon model by performing transfer learning or constructing discriminants from selected output scores.}
\label{fig:sophon}
\end{center}
\end{figure}

In this work, we propose novel LHC experimental methodology called Signature-Oriented Pre-training for Heavy-resonance ObservatioN (abbreviated \textit{Sophon}) to achieve the goal.
This methodology introduces a boosted-jet DNN model (the \textit{Sophon model}) learned from a comprehensive jet dataset.
It is capable of pushing a broad range of hadronic final-state searches toward the sensitivity frontier and also improving model-agnostic approaches.
The Sophon model is pre-trained on a large-scale jet dataset, including various resonance decays that span as wide a range of jet signatures as possible.
Thus, it is expected to learn a comprehensive latent representation of jets.
For the pre-training task, this work implements large-scale classification, using finely categorized labels indicating which partons, leptons, or combinations thereof initiated the jet.
In total, there are 188 classes.
When used in LHC experimental searches, as shown in Fig.~\ref{fig:sophon}, the model offers the ability to construct various tagging discriminants directly from its output nodes, which are already optimized for dedicated signals.
Moreover, one can adopt the transfer learning technique~\cite{survey-tl,survey-deeptl}, specifically, using its latent representation nodes as input to train a lightweight DNN for dedicated model-specific or model-agnostic tasks.
This approach is highly performant in both tagging capabilities and computational efficiency.

The rest of this paper is organized as follows.
Section~\ref{sec:dataset_model} introduces the new dataset, the Sophon model, and training details.
Section~\ref{sec:benchmark} describes a benchmark of its tagging performance. Section~\ref{sec:resonance-search} presents several experiments demonstrating its large potential for LHC resonance searches.
Finally, Section~\ref{sec:outlook} offers our conclusion and outlook.

\section{Dataset and model}
\label{sec:dataset_model}
A prerequisite for making the Sophon method effective is having a large-scale and comprehensive dataset.
We present the \jetclassii dataset, which includes 188 jet classes to facilitate Sophon model training.
Compared with the \jetclass dataset~\cite{pmlr-v162-qu22b,JetClass}, \jetclassii contains resonant jets with a broader mass range and an extended set of final states.
A resonant jet may contain 2, 3, and 4 prongs, where each prong is initiated by a quark, gluon, or lepton.
They are finely categorized with respect to the particle types, quark or lepton flavors, and the tau lepton's decay mode.
Specifically, to generate two-prong jets, we consider a generic spin-0 resonance $X$ with mass up to 500\,GeV, transverse momentum \pt up to 2500\,GeV, either charged or neutral, that decays to diparton $bb$, $cc$, $ss$, $qq$ ($q=u/d$), $bc$, $cs$,  $bq$, $cq$, $sq$, and $gg$, dilepton $ee$, $\mu\mu$, $\tau_{\rm h}\tau_{e}$, $\tau_{\rm h}\tau_{\mu}$, and ditau $\tau_{\rm h}\tau_{\rm h}$ signatures, where $\tau_{e/\mu/{\rm h}}$ represents a tau lepton that subsequently decays into an $e$, $\mu$, or hadrons, respectively.
For jets with 3 or 4 prongs, a decay of $X\to Y^\astinpar Y^\astinpar$ is first performed with a wide $m_Y$ range, $m_Y/m_X\in (0.2,\,0.8)$, followed by a similar diparton and dilepton decay of the secondary resonance $Y$ with the additional inclusion of $Y\to e\nu$, $\mu\nu$, $\tau_{e}\nu$, $\tau_{\mu}\nu$, and $\tau_{\rm h}\nu$ signatures.
Thus, a complete set of jet final states arising from $Y^\astinpar$ pairs is considered.
This results primarily in 4-prong signatures but can also be 3 prongs if an object leaks out of the jet cone or if one of the objects is a neutrino.
In addition to the resonant jet, the background jets from quantum chromodynamics (QCD) multijet events are simulated to cover a wide \pt and mass range, and they are subdivided into 27 classes based on the number of quarks within the jet and their flavors.
A summary of the jet classes can be found in Appendix~\ref{app:jetclassii}.

The simulation of the \jetclassii dataset follows the \jetclass simulation workflow~\cite{pmlr-v162-qu22b}, while additionally emulating for the effect of pileup (PU) with an average of 50 PU interactions and adopting the PU per particle identification algorithm~\cite{Bertolini:2014bba} to remove the PU, with a configuration similar to that used in the CMS experiment~\cite{CMS:2020ebo}.
This creates a more realistic dataset that mimics LHC data collected in Run~2.
Large-$R$ jets are clustered from the processed E-flow objects in the \DELPHES software~\cite{deFavereau:2013fsa} using the anti-\kt algorithm~\cite{Cacciari:2008gp} with $R=0.8$.


To train the Sophon model, 
we use the Particle Transformer (ParT) as the backbone with the same model configuration as in Ref.~\cite{pmlr-v162-qu22b}, except that the fully connected multilayer perceptron (MLP) is now expanded to two layers, first increasing the dimension to 512, then back to $N_{\rm out} = 188$ output nodes.
The neuron values before passing through the MLP have a dimension of $N_{\rm latent}=128$.
They are treated as the Sophon model's latent features used for transfer learning.
Notably, we adopt two special training techniques in addition to the original ParT training~\cite{pmlr-v162-qu22b}.
First, jet samples are selected with a predefined probability during training to ensure a smooth \pt and soft-drop mass (\msd)~\cite{Dasgupta:2013ihk,Larkoski:2014wba} spectrum.
This essentially performs a reweighting on \pt and \msd of the training dataset, a technique previously explored to minimize the tagger's correlation with jet \pt and masses in LHC experiments~\cite{Bradshaw:2019ipy,Moreno:2019bmu,ATL-PHYS-PUB-2023-021}.
The decorrelation with jet mass is especially important for resonance searches to avoid sculpting the background mass distribution when applying a selection on the tagger score.
Secondly, the four-momentum of the jet and its constituents are all scaled by a coefficient to satisfy jet $\pt = 500$\,GeV before gathering the jet inputs. This improves the scale-invariance of the model and generalizes the tagging performance over a wide mass range.
More technical details can be found in Appendix~\ref{app:training}.

\section{Performance benchmark}
\label{sec:benchmark}
We first compare the Sophon model with the best jet taggers achievable in current experiments as a performance benchmark.
Our experiment is delivered on a dataset dedicated to the SM processes, simulated in LHC $pp$ collision at $\sqrt{s} = 13$\,TeV and corresponding to the 100\invfb of data.
This dataset is produced in addition to \jetclassii, using the same configuration for \DELPHES simulation. 
It imposes a trigger requirement dedicated to the boosted topology study, i.e., the scalar \pt sum of all $R=0.8$ jets is larger than 800\,GeV, and one of the jets should have a trimmed mass~\cite{Krohn:2009th} larger than 50\,GeV.
Based on the leading-order cross section calculated from \MGvATNLO~\cite{Alwall:2014hca},
this SM dataset includes $5\times 10^7$ events from QCD multijet process, $9\times 10^5$ $V\,(=W/Z)$+jets events, $3\times 10^5$ events from top quark-antiquark pair ($t\overline{t}$) and single top (ST) quark processes, and other processes including the diboson ($VV$) and Higgs production.

The performance of tagging resonant \xbb and \xbs jets is used for benchmarking.
Here, the \xbb tagging task evaluates the Sophon model's direct tagging capability by constructing discriminants from its output nodes, while \xbs tagging examines its transfer learning ability since there is no direct correspondence in the model's training classes to the $bs$ signature.
The BSM signal process originates from a hypothetical heavy spin-0 (Higgs-boson-like) resonance $X_0$ with a mass equal to 200\,GeV, decaying to $bb$ or $bs$.
For both signal and SM processes, the leading $R=0.8$ jet satisfying the trigger requirement is used for evaluating various algorithms.

For \xbb tagging, we begin by discussing the optimal way to construct discriminants from the Sophon model output. 
A trained multi-class classifier with minimum cross-entropy loss estimates the likelihood ratios of the input classes through the so-called ``likelihood-ratio trick''~\cite{Cranmer:2015bka}.
Specifically, the $i$th ($i=1,\cdots,N_{\rm out}$) classifier output score $g_i(\mathbf{x})$ given input $\mathbf{x}$ satisfies
\begin{equation}\label{eq:multiclass-probs}
    g_i(\mathbf{x}) = \frac{p(\text{class}=i|\mathbf{x})}{\sum_{j=1}^{N_{\rm out}}{p(\text{class}=j|\mathbf{x})}},
\end{equation}
under the ideal DNN assumption, i.e., with sufficient model capacity and data such that the loss reaches the theoretical minimum.
Note that the binary classifier form ($N_{\rm out} = 2$) of this property has been widely explored in high energy physics~\cite{Cranmer:2015bka,Brehmer:2018kdj,Andreassen:2019nnm}, but its extension to multiple classes form has been investigated less.
Here, we show two important properties that can be derived from Eq.~(\ref{eq:multiclass-probs}).
They will guide the construction of Sophon's tagging discriminants throughout this work.
\begin{property}
Class division property: Consider a classifier with an input class $c$ that is subdivided into multiple exclusive subclasses $\{c_1, \ldots, c_N\}$ to form a new classifier.
Let $g_i(\mathbf{x})$ and $g_i'(\mathbf{x})$ denote the output scores of the original and new classifiers, respectively.
The output scores of the two classifiers are related as follows.
\begin{equation}
    g_c(\mathbf{x}) = \sum_{l=1}^{N} g'_{c_l}(\mathbf{x}),\;\; \text{and}\;\; g_i(\mathbf{x}) = g_i'(\mathbf{x})\text{ for }i \neq c.
\end{equation}
\end{property}
\begin{property}
Extraneous classes property: Consider a classifier that is augmented with additional new input classes $\{e_1, \cdots, e_N\}$ to form a new classifier.
The ratios of the output scores for the original classes should remain unchanged, i.e.,
\begin{equation}
    \frac{g_i(\mathbf{x})}{g_j(\mathbf{x})} = \frac{g_i'(\mathbf{x})}{g_j'(\mathbf{x})},\quad \text{for } i,\,j \notin \{e_1, \cdots, e_N\}.
\end{equation}
\end{property}

Built on the above properties, the optimal discriminant for distinguishing \xbb from QCD jets is constructed as
\begin{equation}\label{eq:sophon-xbb-discr}
    \text{discr (\xbb vs. QCD)} = \frac{g_{\xbb}}{g_{\xbb} + \sum_{l=1}^{27} g_{\text{QCD}_l}},
\end{equation}
where $g_{\xbb}$ corresponds the \xbb output score and $g_{\text{QCD}_l}$ corresponds to the scores of 27 QCD classes.
Ideally, this should be equivalent to training a binary classifier DNN to classify the same \xbb jets and the undivided QCD jets, then using the \xbb score as the discriminant.
According to the Neyman--Pearson lemma~\cite{neyman1933ix}, this serves as the strongest discriminant to distinguish the \xbb and QCD jets.

For \xbs tagging, transfer learning is applied to the Sophon model from its latent features with a dimension $N_{\rm latent}=128$, using a two-layer MLP with (512,\,2) nodes.
The parameters of the first linear layer are preloaded from the corresponding part in the Sophon model to ease the learning.
It only has two output nodes for classifying $X\to bs$ jets and QCD jets.
Here, $X\to bs$ jets are again produced with variable $m_X$ in the same kinematics as jets in \jetclassii. The same mass-decorrelation technique is applied during training.
The transfer learning training is much simpler and faster than the original Sophon model training.
Only a small fraction ($1/320$) of the total Sophon training dataset is needed for the transfer learning and the total computational cost (in terms of floating point operations per second) is $1/1\,000\,000$ of the original training.

\begin{figure}[htbp]
\begin{center}
\includegraphics[width=0.48\textwidth]{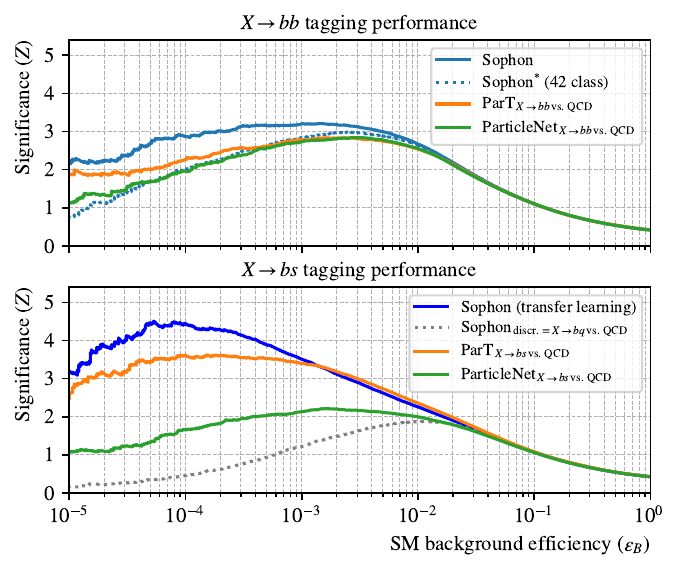}
\caption{Benchmark of Sophon model's performance in the \xbb and \xbs jet tagging tasks,
with the signal originated from a BSM resonance $X_0$ with $m_{X_0}=200$\,GeV that decays to $bb$ and $bs$, and backgrounds corresponding to the 40\invfb of the full SM processes.
The performance of various DNN models is compared in terms of the search's discovery significance versus SM background efficiency calculated within the mass window $150 < \msd < 230$\,GeV.
A major conclusion is that the Sophon model's direct tagging discriminant (for \xbb tagging) and its transfer learning version (for \xbs tagging) both outperform the current best results achievable in the LHC experiment using a ParT or ParticleNet tagger.
It also confirms that the model improves performance when trained by large-scale classification.
}
\label{fig:roc_benchmark}
\end{center}
\end{figure}

Figure~\ref{fig:roc_benchmark} shows the tagging performance in terms of the discovery significance $Z$~\cite{Cowan:2010js} as a function of the SM background selection efficiency, using 40\invfb of data and considering events within the mass window $150 < \msd < 230$\,GeV.
Several tagging models are compared at a given number of signal injections.
To illustrate the current tagging performance achievable at LHC experiments, we train two dedicated tagging models for both tasks: one using state-of-the-art ParT~\cite{pmlr-v162-qu22b} architecture, and one using the ParticleNet model~\cite{Qu:2019gqs}.
These are representative of the current tagging capability within the CMS experiment~\cite{CMS:VHcc,CMS:HH4b,CMS:XY4b}.
These models are trained as binary classifiers to distinguish $X\to bb\,(bs)$ jets against the QCD jets, applying similar training settings.
The performance of the Sophon model in \xbb tagging and its transfer learning version in \xbs tagging already surpasses that of dedicated ParT or ParticleNet trainings.
This demonstrates the ability of the method to adapt to various model-specific jet tagging tasks~\footnote{
Note that the absolute tagging performance does not necessarily match real experiments due to the discrepancies between the \DELPHES modeling and real detector conditions.
Our purpose is to compare methods and draw conclusions about the capabilities of different models.
This will remain valid for real experimental conditions.
}.

Additional comparisons are presented in Fig.~\ref{fig:roc_benchmark}.
First, to study whether the Sophon model gained superior \xbb vs. QCD jet tagging performance from the large-scale classification task, we conduct an ablation study by training the model only on 42 classes (denoted Sophon*), including all 2-prong resonant-jet classes and the QCD classes.
These results show that the Sophon model trained on 188 classes significantly improves the discovery significance at a fixed background efficiency, highlighting the importance of pre-training on a large and comprehensive dataset.
Second, to confirm that the high performance in \xbs vs. QCD jet tagging relies on knowledge transferred from the latent space instead of recycling the tagging ability from existing classification nodes, we identify the output node for \xbq jets that shares the closest similarity with \xbs jets and check the performance when using Sophon's \xbq vs. QCD jet tagging discriminant.
The latter significantly underperforms, confirming the important role of transfer learning.

\section{Implications for resonance search}\label{sec:resonance-search}
After demonstrating the high performance of the Sophon model, we discuss how this approach, once deployed on LHC experiments, will help to accelerate the search for BSM resonances.
We discuss two scenarios to combine the Sophon model with resonance search.

The first method leverages the all-inclusive classification nodes of the Sophon model.
Since we are unsure about the exact final state of the resonant, we can use these 188 scores to make certain combinations, building the numerator and denominator as shown in Fig.~\ref{fig:sophon} (b), to create a discriminant for jet selection.
A typical bump hunt strategy can then be performed on the mass spectrum to search for potential resonances.
This method utilizes the extensive classification ability of the Sophon model to distinguish various jet signatures optimally.
The second method embeds Sophon's transfer learning into fast-evolving model-agnostic search strategies.
Formally, this only involves replacing the existing method's input jet feature space with the Sophon model's latent feature space.
Yet, the extensive knowledge of jet signatures encoded in the feature space is expected to yield improved signal-finding performance for a broad class of signal models.

We evaluate the methods above in the single-jet and the dijet topologies.
The first topology aims to identify resonance structure in a single jet $\msd$ spectrum.
Utilizing the above techniques, we aim to reveal the existence of SM particles amidst the overwhelming QCD multijet backgrounds.
The second experiment performs a standard dijet resonance search to find the resonance peak at the TeV mass scale in the dijet invariant mass \mjj.
This serves as a benchmark for the proposed methods by comparing them with established model-agnostic strategies.

\begin{figure}[htbp]
\begin{center}
\includegraphics[width=0.48\textwidth]{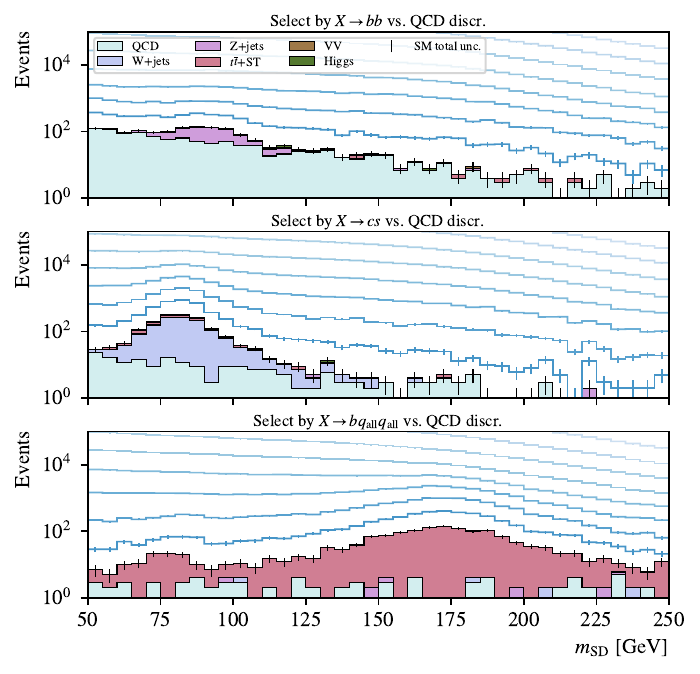}
\caption{Distributions of the leading $R=0.8$ jet \msd for the 40\invfb of simulated SM events by imposing selections on different Sophon tagging discriminants (\xbb, \xcs, and $\xbq_{\rm all}q_{\rm all}$ vs. QCD) at various selection efficiencies $\epsilon$.
The blue and black curves represents $\epsilon$ ranging from $10^{0}$ to $10^{-4}$, where the black one corresponds to $\epsilon=10^{-4}$.
The stacked histograms show the contribution of various SM processes at $\epsilon=10^{-4}$.
The $W/Z/t$ peaks can be revealed from the flat QCD multijet background in different cases.
The graph demonstrates that constructing various tagging discriminants allows signals with corresponding jet signatures to be purified, revealing distinct signal peaks.}
\label{fig:dist_singlejet}
\end{center}
\end{figure}

First, in the single-jet resonance search, we consider the following discriminant to veto QCD jets while purifying certain signal processes,
\begin{equation}
    \text{discr ($A$ vs. QCD)} = \frac{g_{A}}{g_{A} + \sum_{l=1}^{27} g_{\text{QCD}_l}}.
\end{equation}
This study considers three choices for the signature $A$ for illustration:
\begin{align}
\begin{split}
    g_{A_1} &= g_{\xcs}, \qquad g_{A_2} = g_{\xbb}, \\
    g_{A_3} &= g_{\xbq_{\rm all}q_{\rm all}} \equiv \!\!\!\!\!\!\sum_{\text{sig}\in\{ccb, ssb,\atop qqb, bcs, bcq, bsq\}}\!\!\!\!\!\! g_{X\to \text{sig}},
\end{split}
\end{align}
where the $g_{A_3}$ term aims to select all existing 3-prong signatures composed of three quarks with exactly one $b$ quark included.
By selecting jets on the three discriminants, Fig.~\ref{fig:dist_singlejet} shows the change in the \msd spectrum of the 40\invfb of the SM events as the selections become tighter.
The stacked histograms are also shown at a selection efficiency of $\epsilon_B = 10^{-4}$.
Interestingly, the corresponding resonant signatures from the $W$, $Z$ bosons, and the $t$ quark are revealed.
This example demonstrates the Sophon model's broad ability to construct discriminants and sensitively probe resonances with unknown properties.

To show the feasibility of using Sophon's learned knowledge in a model-agnostic search, we use the Simulation Assisted Likelihood-free Anomaly Detection (SALAD) method~\cite{Andreassen:2020nkr} as an illustration.
This method utilizes a simulated background dataset to assist in probing the small number of signal events in the data.
To search for resonance at $\msd \sim m_0$, we define the signal region (SR) as $(m_0-15,\,m_0+15)$\,GeV and the mass sideband (SB) as $(m_0-25,\,m_0-15) \cup (m_0+15,\,m_0+25)$\,GeV.
Intuitively, this method first learns how to reweight the SB simulation to SB data; with this information, it estimates the background density in SR and trains a classifier to distinguish the estimated SR background from the SR data. 
The classifier output is theoretically allowed to identify the signal events in SR optimally.
We choose the QCD background from 20\invfb of data as the simulated background.
For the rest of the dataset, 40\invfb samples are used for training, and the other 40\invfb are used to test the performance.
We apply the method with sliding mass windows, changing $m_0$ from 65 to 295\,GeV with a step of 10\,GeV.
The trained classifier discriminant is applied to the test data in the narrow bin of $(m_0-5,\,m_0+5)$\,GeV at a fixed working point to suppress the QCD backgrounds to the $10^{-3}$ level.
Practically, the major difference compared to the original SALAD method~\cite{Andreassen:2020nkr} is that the input is changed to the jet latent features provided by the Sophon model~\footnote{Note that the original SALAD method is applicable in scenarios where the simulation and data have subtle differences in event generation patterns. This study employs a background simulation using the same generator configuration for simplicity. Nonetheless, It can demonstrate the feasibility of the model-agnostic approach, as the fundamental principle of training a weakly-supervised classifier is preserved under our conditions.}.
\begin{figure}[htbp]
\begin{center}
\includegraphics[width=0.48\textwidth]{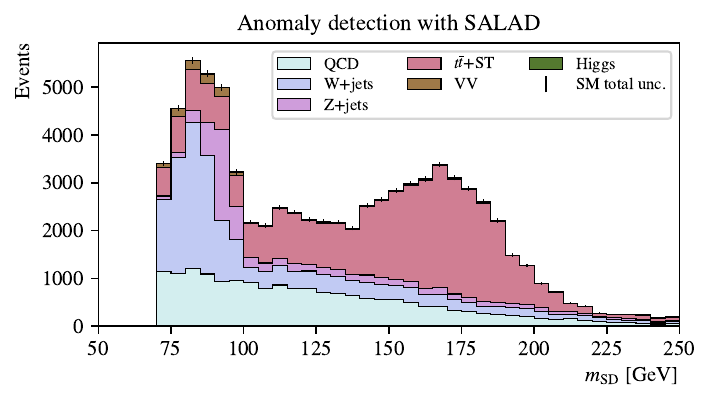}
\caption{Distributions of the leading $R=0.8$ jet \msd for the 40\invfb of simulated SM events by applying the Simulation Assisted Likelihood-free Anomaly Detection (SALAD) method. The classifier is trained with a sliding mass window, and the selection is applied for each classifier at a QCD multijet efficiency working point of $10^{-3}$.
The proportion of non-QCD processes is enhanced. The plot shows the feasibility of combining the transfer learning technique with the model-agnostic search strategy.
}
\label{fig:dist_ad_singlejet}
\end{center}
\end{figure}
Figure~\ref{fig:dist_ad_singlejet} shows the \msd spectrum in the test data after the selection is applied, with the $W/Z/t$ related processes being more pronounced.
Since no dedicated information on the signatures is provided throughout the process, this experiment shows that it is feasible to adopt the Sophon method in model-agnostic searches.

\begin{figure}
\begin{center}
\includegraphics[width=0.48\textwidth]{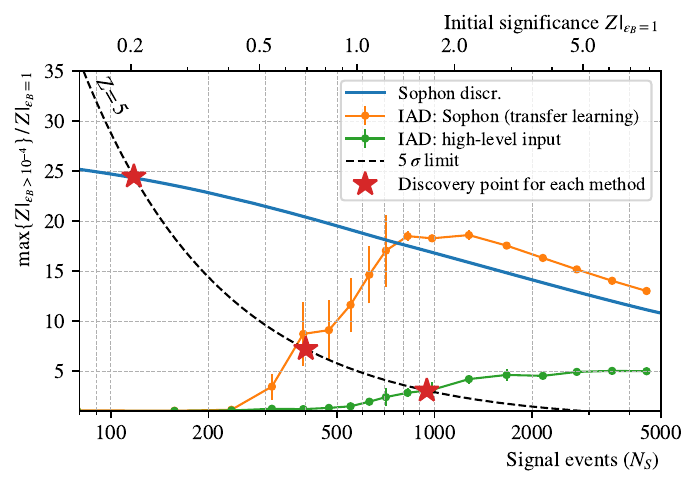}
\caption{Benchmark of the model-agnostic dijet anomaly search capability in search of a triboson BSM signal process from 40\invfb of simulated SM events.
The plot shows the maximum significance improvement defined as $\max\{Z|_{\epsilon_B > 10^{-4}}\} \big/ Z|_{\epsilon_B=1}$ with varying signal injection, within the mass window of the dijet invariant mass $2500 < \mjj < 3100$\,GeV.
The idealized anomaly detection (IAD) limit compares the best performance in different input scenarios, including performing transfer learning on the Sophon model and using high-level jet inputs.
The error bars correspond to the standard deviation of the maximum significance improvement over 20 trainings.
The performance of the constructed Sophon discriminant is also compared.
The $5\,\sigma$-limit curve is highlighted to compare the number of signal events required for each method for establishing a first discovery.
}
\label{fig:ad_dijet}
\end{center}
\end{figure}

We then experiment with the techniques in a widely explored dijet resonance search, aiming to benchmark the proposed methods.
Specifically, we consider a BSM triboson final-state process initiated by $W'\to W\phi \to WWW$, with $m_{W'}=3$~TeV, $m_{\phi}=400$\,GeV and all $W$ boson decaying hadronically. The physics model is discussed in Refs.~\cite{Agashe:2016rle,Agashe:2017wss} and is adopted by the early model-agnostic study~\cite{Collins:2018epr}.
We simulate this physics process with the same simulation workflow as above.
Events passing the trigger must contain at least two $R=0.8$ jets with $\pt > 250$\,GeV and $|\eta|<2.5$.
To search for the resonance $W'$ on the dijet invariant mass \mjj, we define the SR as $\mjj\in (2500,\,3100)$\,GeV and SB as $\mjj\in (2200,\,2500)\cup (3100,\,3400)$\,GeV.
First, we attempt to construct a discriminant with Sophon's output scores to select the expected signature of both jets.
Ideally, we expect one jet to be 2-prong while another jet to be 4-prong; however, given that $m_\phi$ is large, it is probable that the jet only reconstructs three quarks within the cone.
Therefore, we select signatures with either 2 prongs initiated by a $W$ boson, or 3 and 4-prong signatures initiated by $WW$, optimizing an event-selection discriminant as the sum of two jet-tagging discriminates in the form of
\begin{equation}
    \text{event discr.} = \sum_{\text{jet}=1,2} \frac{g_{A,\text{jet}}}{g_{A,\text{jet}} + \sum_{l=1}^{27} g_{\text{QCD}_l,\text{jet}}},
\end{equation}
where $g_{A}$ is defined as
\begin{equation}
    g_{A} \equiv 0.3\,g_{W(2)} + 0.1\,g_{WW(4)} + 0.6\,g_{WW(3)}, 
\end{equation}
and
\begin{align}
\begin{split}
g_{W(2)} &\equiv g_{\xcs} + g_{\xqq}, \\
g_{WW(4)} &\equiv g_{X\to ccss} + g_{X\to qqcs} + g_{X\to qqqq}, \\
g_{WW(3)} &\equiv \!\!\!\!\!\sum_{\text{sig}\in\{ccs, ccq, ssc,\atop ssq, qqc, qqs, qqq\}}\!\!\!\!\! g_{X\to \text{sig}}. \\
\end{split}
\end{align}
This can be treated as a model-specific tagging discriminant dedicated to the triboson phase space.
The model-agnostic search ability based on the weakly-supervised approach is also studied and benchmarked with established strategies.
We experiment with the \textit{idealized} case where the classifier is trained to discriminate data in the SR against the SR backgrounds.
This assumes the SR background is perfectly modeled and thus provides a simple benchmark to set a performance limit for all relevant anomaly detection methods under the same input feature space.
The limit is denoted as the idealized anomaly detection (IAD) limit~\cite{Hallin:2021wme}.
We use the Sophon model's latent features as input to train the IAD classifier, comparing it with using high-level jet inputs adopted by various studies~\cite{Collins:2018epr,Nachman:2020lpy,Andreassen:2020nkr,Stein:2020rou,Amram:2020ykb,Hallin:2021wme,Hallin:2022eoq}.
We evaluate the maximum significance improvement, defined as $\max\{Z|_{\epsilon_B > 10^{-4}}\} \big/ Z|_{\epsilon_B=1}$ when imposing a selection on the IAD classifier discriminant, as a function of the injected signal yield, shown in Fig.~\ref{fig:ad_dijet}.

Our results show that Sophon's transfer learning combined with IAD enhances not only the maximum significance improvement, but also its sensitivity to the signal at a low injection---the initial significance that the method starts to be aware of the existence of signal is around 0.6--1.0\,$\sigma$.
It indicates that by leveraging the learned knowledge of a pre-trained Sophon model, we successfully solve the dilemma that exploring lower-level input features for higher distinguishing capability has to compromise the classifier's sensitivity to low signal injection~\cite{Buhmann:2023acn}~\footnote{On this aspect, a recent work~\cite{Mikuni:2024qsr} in parallel with our study has proposed a similar solution using a pre-trained model for anomaly detection.
Our solution differs from this work by proposing the pre-training dedicated to a multitude of jet signatures and utilizing a more lightweight transfer learning to train the weakly-supervised classifier instead of a full model fine-tuning.}.
Furthermore, we emphasize that a key criterion for evaluating the potential of a method in resonance search is if it can reach the $5\,\sigma$ discovery threshold---widely regarded as the gold standard for new discoveries---in the quickest way.
In this regard, Fig.~\ref{fig:ad_dijet} also highlights the $5\,\sigma$-limit curve and the discovery point for each method, showing that Sophon's IAD method can achieve discovery with 2.4 times fewer required signals compared to the traditional methods using high-level inputs, thanks to both improved classifier performance and enhanced sensitivity at lower signal yields.
On the other hand, Sophon's signal-targeted discriminant enables a much quicker $5\,\sigma$ discovery, as it requires a further 3.5 times fewer signal events than the much-improved IAD method via Sophon's transfer learning.

This finding implies that if we aim to search for resonance signatures composed of fragments initiated from SM particles, the most efficient strategy is simply to construct discriminants in a multitude of forms and then search for a potential resonant peak in each case. It allows us to push the sensitivity of various resonant searches towards its frontier.
On the other hand, for detecting anomalous signals that are totally unlike the known signatures induced by SM particles, the model-agnostic strategy will be advantageous.
As physical priors of signals are recognized as important in model-agnostic searches~\cite{Park:2020pak,Cheng:2024yig}, our approach can conceptually enhance transfer capabilities due to the comprehensive jet phase spaces the Sophon model learns from.
Overall, by exploring both methods in the broad resonance search program at the LHC, we can expect significant potential to improve search sensitivity and, hopefully, accelerate the next possible discovery.

\section{Conclusion and outlook}
\label{sec:outlook}
We propose the \textit{Sophon} methodology for signature-oriented pre-training over a large-scale dataset and preset \jetclassii, which covers comprehensive boosted jet signatures.
Pre-trained on \jetclassii as a large-scale classifier, the \textit{Sophon model} can distinguish over a hundred different jet signatures, showing superior performance in constructed tagging discriminant and transfer learning, outperforming current best results achievable from the LHC experiment.
The resonance search studies suggest that it can push a broad range of resonance searches to the sensitivity frontier and also greatly improve model-agnostic searches.
It opens a promising direction in conducting future boosted-jet searches at the LHC.

Driven by rapid advancements in deep learning for developing large models, recent LHC phenomenology works have focused on jet model pre-training and fine-tuning applications~\cite{Dillon:2021gag,Vigl:2024lat,Heinrich:2024sbg,Birk:2024knn,Mikuni:2024qsr}.
Compared with established studies, our work demonstrates the importance of building large-scale datasets to train an expressive model and highlights its significance for broadly accelerating the resonance search.
Additionally, this methodology can be naturally integrated into the existing LHC analysis workflow. One can store the latent features in the central dataset to facilitate the use of its output scores or the highly efficient transfer learning without the need to revisit the full model.
The inference of the Sophon model is also affordable, as it shares a similar computational cost with the default ParT or ParticleNet architecture~\cite{pmlr-v162-qu22b}.

As this work explores a simple classification approach under the pre-training methodology, further studies may extend the use of jet signatures and combine them into novel training targets to improve the model's expressiveness.
Additionally, exploring novel applications of the Sophon model within LHC data analyses, other than generic resonance searches, can be an interesting task. Addressing challenges such as calibration can pave the way for its broader application.

The \jetclassii dataset (with the \DELPHES configuration) and the Sophon model will be publicly available.


\section*{Acknowledgement}
CL, AA, DF, LG, and QL are supported by National Natural Science Foundation of China (NSFC) under Grants No.~12325504, No.~12061141002, and No.~2075004.
JD and RK are supported by the DOE, Office of Science, Office of High Energy Physics Early Career Research program under Grant No. DE-SC0021187, and the U.S. National Science Foundation (NSF) Harnessing the Data Revolution (HDR) Institute for Accelerating AI Algorithms for Data Driven Discovery (A3D3) under Cooperative Agreement No. PHY-2117997.
JD, GK, and LM acknowledge the support of the Deutsche Forschungsgemeinschaft (DFG, German Research Foundation) under Germany’s Excellence Strategy -- EXC2121 “Quantum Universe” -- 390833306.
The research is supported in part by the computational resource operated at the Institute of High Energy Physics (IHEP) of the Chinese Academy of Sciences.
We appreciate helpful discussions with our colleagues, Anna Benecke, Stephane Cooperstein, Sen Deng, Raffaele Gerosa, Loukas Gouskos, Zichun Hao, Farouk Mokhtar, Spandan Mondal, Alexandre De Moor, Jennifer Ngadiuba, Sitian Qian, Melissa Quinnan, Sebastian Wuchterl, and Yuzhe Zhao in the CMS Collaboration when studying the experimental version of this methodology.
CL thanks Zixun Kou, Haonan Lu, and Fanqiang Meng for discussions with them to prepare the manuscript and for the support from Yannan Pan.

\bibliography{main}

\cleardoublepage

\appendix

\section{Supplementary details on \textsc{JetClass-II}}\label{app:jetclassii}

\textit{Simulation.}---
The \jetclassii dataset includes a variety of resonant jets and QCD jets.
The resonant jets are initiated from (1) $X\to $ 2 prong signatures with neutral resonance $X$, (2) $X\to $ 2 prong signatures with charged resonance $X$, and (3) $X\to Y^\astinpar Y^\astinpar \to$ 4 prong signatures, with an introduction below.

The case (1) is generated by the $p\,p\to H\,H$ process using the \MGvATNLO (MG) 2.9.18~\cite{Alwall:2014hca} generator at the leading order (LO) with the \textsc{heft} model, with 16\,M events in total.
To control the resonant jet's \pt and mass, the minimum \pt of $H$ at the hard-scattering level is sampled in 50 points from $(100,\,2500)$\,GeV in the logarithm spacing and the $H$ mass is uniformly sampled from $(15,\,500)$\,GeV with a step of 5\,GeV.
The decay of the $H$ resonance and parton showering is simulated by \PYTHIA 8.3~\cite{SJOSTRAND2015159}. The decay modes (branching ratio) are $bb~(\frac{1}{8})$, $cc~(\frac{1}{8})$, $ss~(\frac{1}{8})$, $dd~(\frac{1}{16})$, $uu~(\frac{1}{16})$, $gg~(\frac{1}{8})$, $ee~(\frac{1}{16})$, $\mu\mu~(\frac{1}{16})$, $\tau\tau~(\frac{1}{4})$.
The subsequent $\tau$ decay follows the SM configuration.

The case (2) is generated by $p\,p\to H^+\,H^-$ process using MG at LO with the \textsc{2HDM} model, with 12\,M events in total.
The same configuration on the $H^\pm$ minimum \pt and its mass is used.
The decay of the $H^\pm$ resonance and parton showering is simulated by \PYTHIA 8.3. The decay modes (branching ratio) are $du~(\frac{1}{6})$, $su~(\frac{1}{6})$, $bu~(\frac{1}{6})$, $cd~(\frac{1}{6})$, $cs~(\frac{1}{6})$, $bc~(\frac{1}{6})$.

The case (3) is generated by the $p\,p\to h\,h$ process using MG at LO with the \textsc{2HDM} model, with 120\,M events in total.
The same configuration on the $h$ minimum \pt and its mass is used.
The decay processes are simulated by \PYTHIA 8.3, with the resonance $h$ decaying to $HH$ and $H^+H^-$, each with $\frac{1}{2}$ branching ratio. Here, the $h$, $H$, and $H^\pm$ are the two $\mathcal{CP}$-even Higgs and the charged Higgs bosons in the \textsc{2HDM} model.
The mass of $H$ and $H^\pm$ are configured as $\lambda m_h$, where $\lambda$ is sampled uniformly within $(0.2,\,0.8)$ in MG.
$H$ decay is the same with the case (1), while the $H^\pm$ decay is similar to the case (2) but with special inclusion of decay nodes including a neutrino final state. The $H^\pm$ decay modes (branching ratio) are then $du~(\frac{5}{48})$, $su~(\frac{5}{48})$, $bu~(\frac{5}{48})$, $cd~(\frac{5}{48})$, $cs~(\frac{5}{48})$, $bc~(\frac{5}{48})$, $e\nu~(\frac{1}{16})$, $\mu\nu~(\frac{1}{16})$, $\tau\nu~(\frac{1}{4})$.
The subsequent $\tau$ decay follows the SM configuration.

To initiate QCD jets, a $2\to2$ QCD multijet process with the \PYTHIA 8.3 generator is simulated with 20\,M events. The minimum \pt of the hard-scattering process is sampled within $(100,\,5000)$\,GeV in the logarithm spacing to ensure a wider jet \pt and mass coverage.

The simulated events from all the above processes are proceeded with \DELPHES 3~\cite{deFavereau:2013fsa} for fast simulation of the detector response and object reconstruction. The \DELPHES simulation card is adapted from the CMS detector configuration but modifies the impact parameter of charged particles to match with the CMS tracker resolution, similar to the \jetclass simulation~\cite{pmlr-v162-qu22b}
In addition, the pileup (PU) effect with an average of 50 PU interactions are included, adapted from the CMS detector configuration with PU~\cite{deFavereau:2013fsa}.
The PU per particle identification (PUPPI) algorithm~\cite{Bertolini:2014bba} is also applied for PU removal, adapted from the CMS detector configuration in Phase-II~\cite{deFavereau:2013fsa} with additional parameter modifications based on the Phase-I CMS detector condition and taking the CMS experimental configuration as a reference~\cite{CMS:2020ebo}.
The PUPPI algorithm assigns a value between 0 and 1 to each E-flow object that signifies the probability the object originates from the genuine interaction. It scales the object's four-momentum with the value.
The processed E-flow objects are used to cluster large-$R$ jets using the anti-\kt algorithm~\cite{Cacciari:2008gp} with $R_0=0.8$.
Selected jets must satisfy $200 < \pt < 2500$\,GeV, $20 < \msd < 500$\,GeV, and $|\eta| < 2.5$.

\textit{Jet labeling.}---
A large-$R$ jet produced by the diresonant production process is assigned a resonant-jet label if it matches any of the 161 resonant-jet classes or is discarded if it does not match any resonant-jet label.
We count the first-generation decay products of the resonance $X$ or $Y^\astinpar Y^\astinpar$. The $u$ and $d$ quarks are merged with and use $q$ to represent them. The tau lepton is further exclusively divided into three subclasses, $\tau_e$, $\tau_\mu$, and $\tau_{\rm h}$, if it sequentially decays into an electron, muon, or hadrons.
This results in the following truth particle types that may appear in the final state: $b/c/s/q/g/e/\mu/\tau_e/\tau_\mu/\tau_{\rm h}$ (neutrinos should be excluded).
The matching of the jet with a truth label requires the matching rule of all truth particles presented in the label to be satisfied. 
The matching rule is $\Delta R({\text{particle,\,jet}}) < R_0$ for particle types expect for $\tau_e$ or $\tau_\mu$; for the latter, a matching requires their decayed $e$ or $\mu$ daughter satisfies $\Delta R < R_0$ with respect to the jet axis.

A large-$R$ jet produced by the QCD multijet process is exclusively assigned a QCD label according to the number of $b$, $c$, $s$ quarks, read from the \PYTHIA parton list before their hadronization, satisfying $\pt > 10$\,GeV, and matched with the jet axis by $\Delta R < R_0$. For each number, it is categorized into three cases: 0, 1, $\geq 2$. This gives rise to 27 QCD classes in total.

The labels and their indices provided in the \jetclassii dataset are summarised in Table~\ref{tab:jetclassii-labels}.

\begin{table*}
\caption{Summary of the 188 jet labels in the \jetclassii dataset.}
\label{tab:jetclassii-labels}
\footnotesize
\centering
\begin{tabular}{p{0.14\linewidth}p{0.11\linewidth}p{0.72\linewidth}}
\toprule
\textbf{Major types}  & \textbf{Index range} & \textbf{Label names} \\
\midrule
Resonant jets: \newline $X\to$ 2 prong & 0--14 &
$bb$, $cc$, $ss$, $qq$, $bc$, $cs$, $bq$, $cq$, $sq$, $gg$, $ee$, $\mu\mu$, $\tau_{\rm h}\tau_{e}$, $\tau_{\rm h}\tau_{\mu}$, $\tau_{\rm h}\tau_{\rm h}$\\
\midrule
Resonant jets: \newline $X\to$ 3 or 4 prong & 15--160 &
$bbbb$, $bbcc$, $bbss$, $bbqq$, $bbgg$, $bbee$, $bb\mu\mu$,
$bb\tau_{\rm h}\tau_{e}$, $bb\tau_{\rm h}\tau_{\mu}$, $bb\tau_{\rm h}\tau_{\rm h}$,
$bbb$, $bbc$, $bbs$, $bbq$, $bbg$, $bbe$, $bb\mu$,
$cccc$, $ccss$, $ccqq$, $ccgg$, $ccee$, $cc\mu\mu$,
$cc\tau_{\rm h}\tau_{e}$, $cc\tau_{\rm h}\tau_{\mu}$, $cc\tau_{\rm h}\tau_{\rm h}$,
$ccb$, $ccc$, $ccs$, $ccq$, $ccg$, $cce$, $cc\mu$,
$ssss$, $ssqq$, $ssgg$, $ssee$, $ss\mu\mu$,
$ss\tau_{\rm h}\tau_{e}$, $ss\tau_{\rm h}\tau_{\mu}$, $ss\tau_{\rm h}\tau_{\rm h}$,
$ssb$, $ssc$, $sss$, $ssq$, $ssg$, $sse$, $ss\mu$,
$qqqq$, $qqgg$, $qqee$, $qq\mu\mu$,
$qq\tau_{\rm h}\tau_{e}$, $qq\tau_{\rm h}\tau_{\mu}$, $qq\tau_{\rm h}\tau_{\rm h}$,
$qqb$, $qqc$, $qqs$, $qqq$, $qqg$, $qqe$, $qq\mu$,
$gggg$, $ggee$, $gg\mu\mu$,
$gg\tau_{\rm h}\tau_{e}$, $gg\tau_{\rm h}\tau_{\mu}$, $gg\tau_{\rm h}\tau_{\rm h}$,
$ggb$, $ggc$, $ggs$, $ggq$, $ggg$, $gge$, $gg\mu$,
$bee$, $cee$, $see$, $qee$, $gee$,
$b\mu\mu$, $c\mu\mu$, $s\mu\mu$, $q\mu\mu$, $g\mu\mu$,
$b\tau_{\rm h}\tau_{e}$, $c\tau_{\rm h}\tau_{e}$, $s\tau_{\rm h}\tau_{e}$, $q\tau_{\rm h}\tau_{e}$, $g\tau_{\rm h}\tau_{e}$,
$b\tau_{\rm h}\tau_{\mu}$, $c\tau_{\rm h}\tau_{\mu}$, $s\tau_{\rm h}\tau_{\mu}$, $q\tau_{\rm h}\tau_{\mu}$, $g\tau_{\rm h}\tau_{\mu}$,
$b\tau_{\rm h}\tau_{\rm h}$, $c\tau_{\rm h}\tau_{\rm h}$, $s\tau_{\rm h}\tau_{\rm h}$, $q\tau_{\rm h}\tau_{\rm h}$, $g\tau_{\rm h}\tau_{\rm h}$,
$qqqb$, $qqqc$, $qqqs$,
$bbcq$,
$ccbs$, $ccbq$, $ccsq$,
$sscq$,
$qqbc$, $qqbs$, $qqcs$,
$bcsq$,
$bcs$, $bcq$, $bsq$, $csq$, 
$bce\nu$, $cse\nu$, $bqe\nu$, $cqe\nu$, $sqe\nu$, $qqe\nu$,
$bc\mu\nu$, $cs\mu\nu$, $bq\mu\nu$, $cq\mu\nu$, $sq\mu\nu$, $qq\mu\nu$,
$bc\tau_{e}\nu$, $cs\tau_{e}\nu$, $bq\tau_{e}\nu$, $cq\tau_{e}\nu$, $sq\tau_{e}\nu$, $qq\tau_{e}\nu$,
$bc\tau_{\mu}\nu$, $cs\tau_{\mu}\nu$, $bq\tau_{\mu}\nu$, $cq\tau_{\mu}\nu$, $sq\tau_{\mu}\nu$, $qq\tau_{\mu}\nu$,
$bc\tau_{\rm h}\nu$, $cs\tau_{\rm h}\nu$, $bq\tau_{\rm h}\nu$, $cq\tau_{\rm h}\nu$, $sq\tau_{\rm h}\nu$, $qq\tau_{\rm h}\nu$ \\
\midrule
QCD jets & 161--187 & $bbccss$, $bbccs$, $bbcc$, $bbcss$, $bbcs$, $bbc$, $bbss$, $bbs$, $bb$,
$bccss$, $bccs$, $bcc$, $bcss$, $bcs$, $bc$, $bss$, $bs$, $b$,
$ccss$, $ccs$, $cc$, $css$, $cs$, $c$, $ss$, $s$, others \\
\bottomrule
\end{tabular}
\end{table*}

The total number of the labeled jets in the \jetclassii dataset is around 139\,M.
This includes 22\,M resonant jets with 2 prongs, 99\,M resonant jets with 3 or 4 prongs, and 18\,M QCD jets.

\textit{Jet variables.}---
The jet constituent features are used as the input for neural network training.
These features are carried on E-flow objects and include kinematic features, particle identification variables, and impact parameters features, closely following the \jetclass dataset.
The jet kinematic variables are also provided in the dataset used to construct the neural work input.

Additional variables include the jet $N$-subjettiness variables~\cite{Thaler:2010tr} up to $N=4$ as a representative of the high-level jet observables, and several generator-level variables indicating the jet signatures.
These include the jet label, assigned by counting for the matched truth particles introduced above, and a list of the matched particles with their type and kinematic features included.

\section{Supplementary details on trained models}\label{app:training}

\subsection{Sophon model}
The Sophon model adopts the Particle Transformer architecture following Ref.~\cite{pmlr-v162-qu22b} with the fully connected multilayer perception (MLP) extended to a two layers with dimensions of (512,\,188).
The main body of the Sophon model includes 6 particle attention blocks and 2 class attention blocks, with an embedding dimension of 128, and the number of heads equals 8.
The initial particle features are embedded with a 3-layer MLP with (128,\,512,\,128) nodes, and the pairwise particle features are embedded with a 4-layer elementwise MLP with (64,\,64,\,64,\,8) nodes. The GELU nonlinearity is used throughout the model. The Sophon model includes 2.3\,M parameters in total.

The model takes input from all jet constituents, including the kinematic features, particle identification, and impact parameter features.
It adopted the scaled kinematics inputs, where features related to the constituent or jet four-momentum are all scaled by a parameter such that the jet \pt after scaling is 500\,GeV.

The procedure of sampling-based reweighting from the training dataset to decorrelate the tagger score with jet \pt and \msd is introduced as follows.
The training samples are selected into the training pool with a predefined probability during the on-the-fly data loading process.
These probabilities serve as reweighting factors that reweight the two-dimensional histograms bin by bin, constructed by jet \pt and \msd within the range of $200 < \pt < 2500$\,GeV and $20 < \msd < 500$\,GeV. 
The target is to yield the same normalized distributions for several specific reweighting classes.
The reweighting classes are formed by merging 188 finely classified categories to some extent: classes with only quark flavor differences have been merged, and all 27 QCD jet classes have been merged into one.
This results in 30 reweighting classes.
In addition, the relative weights of the 30 reweighting classes are properly chosen to weigh the number of samples in the training pool for each classes.

The model is trained with a batch size of 512 with an initial learning rate (LR) of $5\times 10^{-4}$. 
The full \jetclassii dataset is split into 80/20\% for each file to serve as the training/validation set.
It is trained over 80 epochs, with each epoch iterating over 10\,M samples.
The optimizer and the LR scheduler are the same as the ParT training~\cite{pmlr-v162-qu22b}.
We use the \textsc{Lookahead} optimizer~\cite{zhang2019lookahead} with $k = 6$ and $\alpha = 0.5$ and the inner optimizer is \textsc{RAdam}~\cite{liu2019variance} with $\beta_1 = 0.95$, $\beta_2 = 0.999$, and $\epsilon = 10^{-5}$. 
The LR remains constant for the first 70\% of the iterations, and then decays exponentially, changing at the beginning of every following epoch, down to 1\% of the initial value at the end of the training.
A model checkpoint is saved in every epoch, and the checkpoint with the highest
accuracy on the validation set is chosen.

\subsection{Sophon model* (42-class)}
This model adopted the same Sophon model configuration except that the classification node dimension is modified to 42.
It is trained to classify a subset of jet signatures, which covers all the final states from 2-prong resonant jets and the QCD jets.
The training dataset then corresponds to the 2-prong resonant and QCD jets, summed up to 40\,M jets.

Compared to the Sophon model training, it is trained over 80 epochs, with each epoch iterating 2.5\,M samples. The other training configurations are the same as the Sophon model case.

\subsection{ParT model for \xbb $(bs)$ vs. QCD}

This binary classifier with the ParT architecture~\cite{pmlr-v162-qu22b} is used
to benchmark the current state-of-the-art discrimination capability between the resonant \xbb $(bs)$ jets and QCD jets.
It adopts the same ParT model configuration of Ref.~\cite{pmlr-v162-qu22b} with two classification nodes.
The input features use the original ParT training input, i.e., no scaling of the jet and constituent four-momenta is applied. This is found to have a negligible difference compared to the scaling case when evaluating the \xbb $(bs)$ vs. QCD tasks.

The training is performed on the labeled \xbb $(bs)$ jets and the QCD jets.
To prevent too few signal \xbb $(bs)$ jets in the \jetclassii dataset which might limit the model's capabilities,
we produce an extended size of the \xbb $(bs)$ jets with the same configuration to enlarge this dedicated category, amounting to 16\,M jets each, reaching a similar size with the QCD jets.
Also, a similar sampling-based reweighting is performed to reach the same 2D histogram on jet \pt and \msd for the two classes, with their relative class weights set as $1:1$.

Compared to the Sophon model training, it is trained over 50 epochs, with each epoch iterating 2.5\,M samples.
A weight decay of 0.01 is adopted to achieve better performance. The other training configurations are the same as above.

\subsection{ParticleNet model for \xbb $(bs)$ vs. QCD}

The binary classifier with the ParticleNet architecture~\cite{Qu:2019gqs} is used to represent the tagging capability achievable in several present CMS analyses.
The ParticleNet model adopts the same configuration as in Ref.~\cite{Qu:2019gqs}.
It comprises three edge convolution blocks with increased dimensions of (64,\,128,\,256) with the number of nearest neighbors for edge computing set as 16.
The input to ParticleNet is similar to the experiment on \jetclass shown in Ref.~\cite{pmlr-v162-qu22b}.
The training is performed on the same extended signal \xbb $(bs)$ jets, and the QCD jets in \jetclassii. The same sampling-based reweighting on the two classes is adopted as introduced above.

The ParticleNet classifier is trained over 50 epochs, with each epoch iterating 2.5\,M samples. A batch size of 512 and initial LR of $5\times 10^{-3}$ is adopted. No weight decay is applied in the ParticleNet training. The optimizer and the LR scheduler are the same as above.

\subsection{Transfered Sophon model for \xbs vs. QCD}

To transfer the knowledge of the Sophon model to perform \xbs vs. QCD jet tagging task, latent space features with a dimension of 128 are used as input to train a 2-layer MLP with (512,\,2) nodes with \textsc{ReLU} nonlinearity. Parameters of the first layer $\texttt{Linear}(188,\,512)$ are preloaded from the corresponding layer in the original Sophon model.
The training is performed on the same extended \xbs resonant jet dataset, and the QCD jet from \jetclassii. The same sampling-based reweighting on the two classes is used.

The training is performed in 1 epoch, which iterates only 2.5\,M samples to reach a convergence. A batch size of 1024 and a constant LR of $5\times 10^{-4}$ is adopted.

\subsection{SALAD classifiers for single-jet resonance search}

In the generic search of resonances on the leading jet \msd spectrum, the Simulation Assisted Likelihood-free Anomaly Detection (SALAD) method~\cite{Andreassen:2020nkr} is used, involving training two classifiers.
The classifier is combined with the Sophon's transfer learning concept; hence, the input of the classifier is the dimensional 128 latent features of the leading jet provided by the Sophon model.

The first classifier is trained to distinguish the QCD background and all events (including QCD jets and other SM processes), in the mass sidebands (SBs) $\msd \in (m_0-25,\,m_0-15)\cup (m_0+15,\,m_0+25)$\,GeV at a given $m_0$. The sampling rate from the two classes is controlled by the data loader to yield the same number of events.
The classifier network is a 3-layer MLP with (512,\,64,\,2) nodes and \textsc{ReLU} nonlinearity, where a preloading of parameters of the first layer is also done.
The training is performed in 1 epoch to iterate 5\,M samples. It takes a batch size of 50\,000 and a constant LR of $1\times 10^{-3}$.
The score corresponding to the all-event class $w(\mathbf{x})$ predicted by the network is used in the next step.
Notably, An ensemble of 100 networks is trained, and the averaged score $\bar{w}(\mathbf{x})$ is used.

The second classifier is trained to distinguish the QCD background and all events in the signal regions (SRs) $\msd \in (m_0-15,\,m_0+15)$\,GeV.
The training uses the same MLP configuration and the preloading of parameters.
It also iterate 5\,M samples and takes a batch size of 50\,000 and a constant LR of $1\times 10^{-3}$.
As proposed in the SALAD method, the per-sample loss function is multiplied by $\bar{w}(\mathbf{x})/(1 - \bar{w}(\mathbf{x}))$ if it belongs to the QCD class.
Similarly, an ensemble of 100 networks is trained, and an average of scores that corresponding to the all-event node is used as the final discriminant to suppress the backgrounds.

\subsection{IAD classifiers for dijet resonance search}

In the dijet resonance search experiment, we use 40\invfb SM data to train the classifier and 40\invfb data for test.
This amounts to around 330\,k SM events in SR $\mjj\in (2500,\,3100)$\,GeV.
The former is further split to 80/20\% for training/validation.

We train two kinds of classifiers under the idealized anomaly detection (IAD)~\cite{Hallin:2021wme} scheme that can represent the sensitivity boundary of the weakly-supervised methods to detect anomalous resonance.
The classifiers differ by their input features.
In the IAD scheme, the classifier is trained to distinguish the backgrounds and the data events in SR, where the backgrounds correspond to all SM processes, and the data additionally includes the injected signal events.
The experiment evaluates the classifier performance as a function of the injected signal $N_s$ in SR.
Practically, for each $N_s$, we divide the 330\,k SM events into two parts and inject $N_s/2$ signal events in one part. The classifier is trained to distinguish the two parts.
For each $N_s$, the experiments are repeated 100 times with different $N_s$ signals chosen for the study. This allows us to calculate the mean and standard deviation of the classifier performance under each $N_s$.

The first classifier is an application of Sophon's transfer learning. We use a 3-layer MLP with (512,\,64,\,2) nodes and \textsc{ReLU} nonlinearity. Here, the input is the two latent feature vectors with dimension 128 from the two jets.
To proceed with the input with two vectors to the network, the vectors are first passed through the first and second layers of the MLP, respectively. The resulting outputs are then summed and passed through the third layer.
An ensemble of 100 networks is trained, and their averaged score corresponding to the data node is used as the discriminant.


The second classifier applies to the high-level input from the dijet system following Ref.~\cite{Collins:2018epr}. For each jet, we consider the jet \msd, the number of constituent $N_{\rm const.}$, the $N$-subjettiness variable~\cite{Thaler:2010tr} $\tau_1$, $\tau_2$, $\tau_3$, and $\tau_4$, taking the logarithm of each variable and standardize it within the range between $-2$ and 2.
The input takes two dimension-6 vectors for both jets.
We use a 4-layer MLP with (512,\,128,\,128,\,2) nodes and \textsc{ReLU} nonlinearity.
Similarly, the vectors are first passed through the first and second layers of the MLP, respectively, and then the outputs are summed and passed through the third and fourth layers.
An ensemble of 100 networks is trained, and their averaged score corresponding to the data node is used as the discriminant.

\end{document}